# Generalized formula for the Landau-Zener transition in interacting Bose-Einstein condensates

A.M. Ishkhanyan

*Institute for Physical Research NAS of Armenia, 0203 Ashtarak-2, Armenia*

E-mail: aishkhanyan@gmail.com

**Abstract.** We present a rigorous analysis of the generalized Landau-Zener problem for the two-level interacting Bose-Einstein condensates. We show that the dynamics of the system is accurately, in detail, described by a two-term variational ansatz that is valid for the whole time domain and is applicable for any set of involved parameters. Applying an exact third order nonlinear differential equation we construct an advanced fifth order polynomial equation for the final transition probability serving as a highly accurate generalized Landau-Zener formula.

The Landau-Zener-Stuckelberg-Majorana linear-in-time term-crossing problem [1, 2, 3, 4] is a well appreciated basic model that serves as a prototype for all level crossing quantum models. Since the realization of the Bose-Einstein condensates in dilute gases of neutral atoms [5, 6] different nonlinear generalizations of this problem have been suggested and explored both theoretically and experimentally [7-22]. We present a comprehensive analysis of the problem on the basis of a two-term variational ansatz that accurately, in detail, describes the time evolution of the system in the whole time domain. A remarkable result of the present development is a highly accurate generalized formula for the final Landau-Zener transition probability which is applicable for the whole variation range of the involved parameters.

The time-dependent mean-field two-level problem involving third-order nonlinearities we consider is defined by the following set of coupled first-order nonlinear equations [7-11]

$$i\frac{da_1}{dt} = U(t)e^{-i\delta(t)}a_2 + (\Lambda_{11}|a_1|^2 + \Lambda_{12}|a_2|^2)a_1,$$

$$i\frac{da_2}{dt} = U(t)e^{i\delta(t)}a_1 + (\Lambda_{21}|a_1|^2 + \Lambda_{22}|a_2|^2)a_2,$$ (1)

where $t$ is the time, $a_1, a_2$ are the fist and second states' probability amplitudes, respectively, $U(t)$ is referred to as the Rabi frequency of the laser field, $\delta(t)$ is the corresponding frequency detuning modulation function, and the cubic nonlinearities describing the interparticle elastic interactions are characterized by (real) coefficients $\Lambda_{jk}$. The Landau-Zener field configuration $\{U(t), \delta(t)\}$ we discuss is the one for which the detuning is assumed to be a linear function of time: $\dot{\delta}(t) = 2\delta_0 t$, and the Rabi frequency is supposed to be constant: $U(t) = U_0 = \text{const}$. We treat the basic case when the evolution of the system starts from the first level, i.e., the initial conditions imposed are $a_1(-\infty) = 1$ and $a_2(-\infty) = 0$.

We start by noting that system (1) describes a lossless process so that the total number of particles is conserved: $|a_1|^2 + |a_2|^2 = \text{const} = 1$. Taking into account this motion integral, it can be shown that the transition probability $p = |a_2|^2$ satisfies the following dimensionless nonlinear ordinary differential equation of the third order:

$$\left(\frac{d}{dt'} - \frac{1}{G}\frac{dG}{dt'}\right)\left(\frac{d^2 p}{dt'^2} - \frac{\lambda}{2}(4-8p)\right) + G^2 \frac{dp}{dt'} = 0, \quad (2)$$

where
$$G = 2t' - \Lambda_a + 2\Lambda_s p, \quad (3)$$

$$\Lambda_a = \frac{\Lambda_{11} - \Lambda_{12}}{\sqrt{\delta_0}}, \quad \Lambda_s = \frac{\Lambda_{11} - 2\Lambda_{12} + \Lambda_{22}}{2\sqrt{\delta_0}} \quad (\Lambda_{12} = \Lambda_{21}), \quad (4)$$

$t' = \sqrt{\delta_0} t$ is the dimensionless time and $\lambda$ is the conventional (dimensionless) Landau-Zener parameter: $\lambda = U_0^2 / \delta_0$. For convenience, below we omit the prime for the time variable.

This equation suggests a few immediate observations. First, it is seen that if the cubic nonlinearities are not taken into account the function $G(t)$ coincides with the Landau-Zener detuning. Hence, this function plays the role of an effective (nonlinear) detuning and the point $t = t_{res}$ defined from the condition $G(t_{res}) = 0$ is the point of the effective resonance crossing. Thus, we conclude that the introduction of the cubic nonlinearities results in a *nonlinear shift of the resonance position*. Next, the structure of the effective detuning $G$ suggests that at sufficiently large absolute values of the time variable $t$, when the condition $|2t'| \gg |\Lambda_a - 2\Lambda_s p|$ holds, the role of the nonlinear terms proportional to the interaction parameters $\Lambda_{jk}$ becomes negligible if the time dynamics at this region is considered.

Further, we notice that the parameter $\Lambda_a$ merely leads to a constant shift in the detuning which can be eliminated by the change of the time variable $t_1 = t - \Lambda_a/2$. This change does not affect the initial conditions (since they are imposed at infinity $t = -\infty$) as well as the final transition probability at $t \to +\infty$. For simplicity of notation, below we omit the index 1 (this is formally equivalent to putting $\Lambda_a = 0$). Thus, we put $\Lambda_a = 0$ and the interparticle interaction is now described by a sole combined parameter $\Lambda_s$. As it is seen from Eq. (2), there exist nonzero parameters $\Lambda_{jk}$ for which the interparticle interactions result only in the shift of the detuning by a constant which is eliminated by the mentioned change of the time variable. Of course, this occurs when the parameter $\Lambda_s$ is equal to zero.

Based upon our previous experience (see, e.g., [23, 24, 25]), to describe the time behavior $p = p(t)$, we introduce the following *two-term ansatz*

$$p = p_0(t) + C_1 \frac{p_{LZ}(\lambda_1, t - t_1)}{p_{LZ}(\lambda_1, \infty)}. \tag{5}$$

Here, $C_1$ is a scaling parameter, $t_1$ is a time-shift parameter, $p_{LZ}(\lambda_1, t)$ is the solution of the *linear* Landau-Zener problem for an effective Landau-Zener parameter $\lambda_1$:

$$\left(\frac{d}{dt} - \frac{1}{t}\right)\left(\frac{d^2 p_{LZ}}{dt^2} + 4\lambda_1 p_{LZ} - 2\lambda_1\right) + 4t^2 \frac{d p_{LZ}}{dt} = 0, \tag{6}$$

and $p_0(t)$ is the solution of a nonlinear first-order *augmented limit equation* [23] (both $p_{LZ}(\lambda_1, t)$ and $p_0(t)$ are supposed to satisfy the initial condition $p(-\infty) = 0$). The latter equation is constructed trying to replace in Eq. (2) the second derivative $d^2 p/dt^2$ by a function that resembles the essential features of this term but for which the solution to the equation can be derived. It has previously been shown that in the case of the quadratic-nonlinear Landau-Zener problem the simplest choice of this replacement by a constant term $A$ leads to highly accurate results [23]. The basic idea here is to justify the introduced constant in a way to cancel the mean action of the terms controlled by the logarithmic derivative $G_t/G$. To proceed with this idea, one should look at the most effective time point. It is readily understood from physical point of view and it is immediately seen mathematically that this point is the resonance crossing point $t_{res}$ because here the mentioned derivative diverges. It turns out that in the case considered in the present paper more productive is the replacement $d^2 p/dt^2 \to A_0 + A_1 p$. Indeed, the numerical simulations reveal that one can always find specific values of the variational parameters $A_0, A_1$ such that the function (5) accurately fits the numerical solution to the exact equation (2) in the whole time domain.

The resultant limit first order differential equation is readily integrated to produce a *quartic* polynomial equation for limit function $p_0(t)$ with variational parameters $\alpha_1$ and $\beta_1$ (originating from $A_0$ and $A_1$):

$$\frac{\lambda}{G^2(p_0)} = -\frac{1}{4}\frac{p_0(p_0-\beta_1)+C_0}{(p_0-\alpha_1)^2}. \tag{7}$$

Note that for the initial condition $p(-\infty)=0$ the integration constant $C_0$ vanishes: $C_0=0$. Further, the meaning of involved variational parameters $\alpha_1$ and $\beta_1$ is readily understood when examining the limits $t \to t_{res}$ and $t \to +\infty$. Since in the first case the left-hand side of Eq. (7) diverges, it follows that at this point $p_0$ should adopt value $\alpha_1$, i.e., $p_0(t_{res})=\alpha_1$, so that $\alpha_1$ is the transition probability at resonance crossing point $t=t_{res}$ given by the limit function $p_0$. In the second limit, at $t \to +\infty$, the left-hand side of the equation vanishes, hence, should be $p_0(+\infty)=\beta_1$, thus, $\beta_1$ defines the final transition probability achieved by the limit function $p_0(t)$ (Fig. 1). It is readily understood that in the linear limit $\Lambda_s \to 0$ the function $p_0$ should vanish, hence, at this limit both $\alpha_1$ and $\beta_1$ tend to zero.

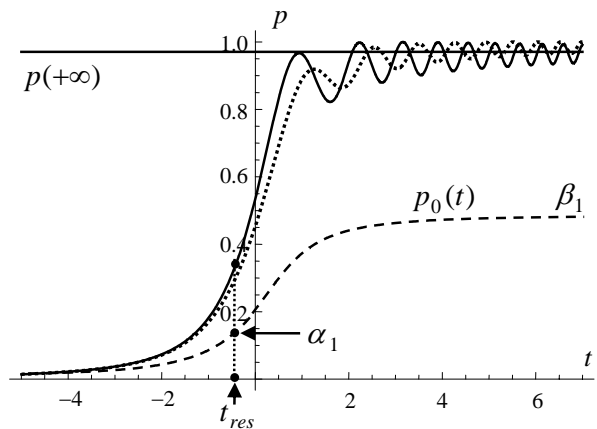

Fig. 1. Nonlinear (solid line) and linear (dotted line) transition probabilities as a function of time at $\lambda=1.5$ and $\Lambda_s=0.5$ (horizontal solid line is the final transition probability $p(+\infty)$). Dashed line shows the limit solution $p_0$ defined by Eq. (7). The position of the resonance crossing $t=t_{res}$ is marked by filled circles.

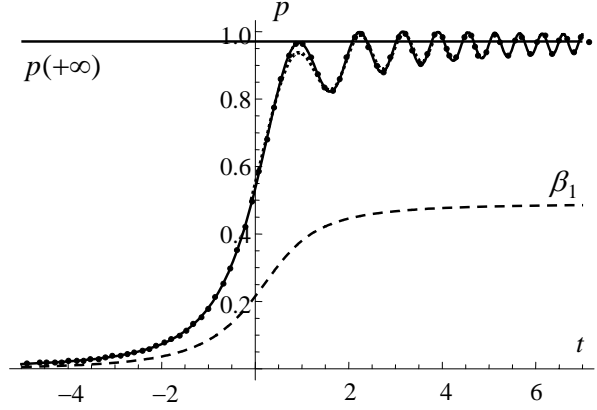

Fig. 2. Nonlinear transition probability $p$ as a function of time at $\lambda = 1.5$ and $\Lambda_s = 0.5$. Dotted line presents the approximation by ansatz (5) with limit Eq. (7), dashed line is the limit solution $p_0$, horizontal solid line is the final transition probability $p(+\infty)$, the filled circles show the time points used for fit. A slight deviation from the exact numerical result (solid line) is observed only in a small region embracing a few first oscillation periods.

The ansatz (5) with Eq. (7) provides a highly accurate approximation if $|\Lambda_s| < U_0$. An example of the fit produced by this formula is shown in Fig. 2. It is seen that the suggested analytic solution is practically indistinguishable from the exact numerical result almost in the whole time variation domain. A slight deviation is observed only in a small region embracing a few first oscillation periods after the system passes the resonance, though even here the deviation is only on the level of a few percents. In order to better examine the substantial progress achieved by the presented development we note the drastic difference of the corresponding nonlinear and linear solutions (Fig. 1).

Discussing now the final transition probability $p_{\inf}$ at $t \to +\infty$ we note that a much better fit is achieved when approximating the second derivative term $d^2 p / dt^2$ in Eq. (2) by a *quadratic* polynomial in $p$. In this case we derive a *fifth-order* limit equation of the form

$$\frac{\lambda}{G^2(p_0)} = \frac{k^2}{9} \frac{(p_0 - \beta_0)(p_0 - \beta_1)(p_0 - \beta_2)}{(p_0 - \alpha_1)^2 (p_0 - \alpha_2)^2}, \quad (8)$$

where for definiteness, without loss of generality, we adopt the numeration of the involved parameters as $\alpha_1 < \alpha_2$ and $\beta_0 < \beta_1 < \beta_2$. This equation suggests several developments. A notable result is achieved when applying a limiting procedure involving simultaneous limits $t \to +\infty$ and $\delta_0 \to 0$ in a way that the left-hand side of Eq. (8) becomes $\lambda/(2\Lambda_s p_{\inf})^2$. Note that the dimensionless quantity $\lambda/(2\Lambda_s p_{inf})^2 \sim U_0^2/(\Lambda_{11} - 2\Lambda_{12} + \Lambda_{22})^2$ does not depend on $\delta_0$. Of particular interest is now the linear limit $\Lambda_s = 0$. Since in this case the left-hand side of the equation diverges it is clear that at this point the parameter $\alpha_1$ adopts the ordinary linear Landau-Zener transition probability $P_{LZ}(\lambda) = 1 - e^{-\pi\lambda}$, i.e., $\alpha_1(\Lambda_s = 0) = P_{LZ}$. In the same way, considering the limits $\Lambda_s \to \pm\infty$ we conclude that $\beta_0 = 0$ and $\beta_1 = 1$. The resulting equation for $p_{inf}(\lambda, \Lambda_s)$ thus reads

$$\frac{9\lambda}{(2k\Lambda_s p_{inf})^2} = \frac{p_{inf}(p_{inf} - 1)(p_{inf} - \beta_2)}{(p_{inf} - P_{LZ})^2(p_{inf} - \alpha_2)^2}. \tag{9}$$

Numerical testing shows that this equation accurately describes the final transition probability for all the variation range of input parameters of the problem under consideration, $U_0$, $\delta_0$ and $\Lambda_{jk}$. The accuracy of the suggested approximation is demonstrated in Figs. 3.1-3.2 for several values of the Landau-Zener parameter $\lambda$ ($\lambda = 10, 1, 0.2, 0.05$). We note that the description is very good for $\lambda \geq 1$ (Fig. 3.1). Furthermore, it is seen that for $\lambda < 1$ the coincidence is good almost everywhere. Except a small region of negative $\Lambda_s$ (Figs. 3.2), the absolute error is of the order of $10^{-4}$. Note that for this region of negative $\Lambda_s$ and relatively small $\lambda$ the deviation is also rather small. A notable observation is that for small $\lambda \ll 1$ and negative $\Lambda_s$ the final transition probability undergoes abrupt, nearly vertical jump to $p_{inf} = 1$ followed by expressed oscillations which dump to the line $p_{inf} = 1$ (Figs. 3.2a, 3.2b). This feature is not described by the suggested equation (9). However, we have checked that a modification of the proposed approach is potent to describe this peculiarity as well. We will address this point in a separate publication because these oscillations suggest a different physical behavior that deserves a separate treatment.

Finally, we numerically find that the variational parameters $k$, $\alpha_2$ and $\beta_2$ are approximated as

$$k(\lambda) = 1 - e^{-\lambda}, \quad \alpha_2 = 2.5, \quad \beta_2(\lambda) = 1 + \frac{12}{(1+\lambda)^3} \tag{10}$$

(the accuracy of the last formula is shown on Fig. 4).

These relations together with Eq. (9) fulfill the development. The derived equation (9) can be viewed as a generalized Landau-Zener formula for the two-level interacting Bose-Einstein condensates. For completeness, on Fig. 5 we show the family of curves $p_{inf}(\Lambda_s)$ defined by this formula for several $\lambda$.

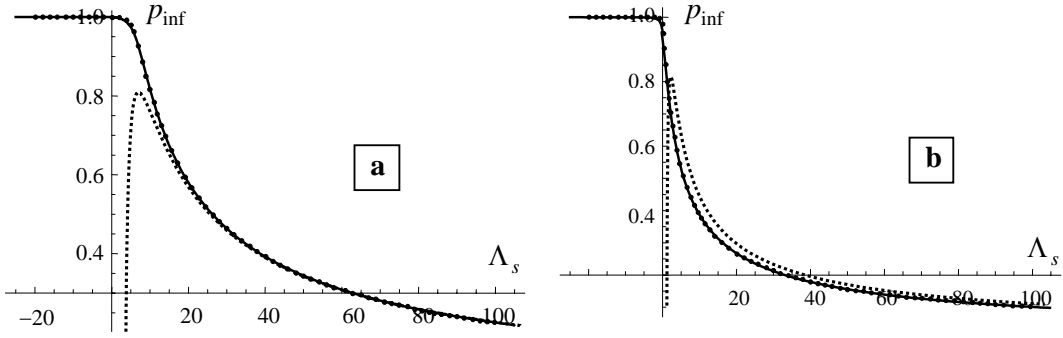

Fig. 3.1. Final transition probability $p_{inf}$ as a function of $\Lambda_s$ (solid lines - Eq. (9), filled circles – numerical result, dotted lines - Eq. (11)): (a) $\lambda = 10$ and (b) $\lambda = 1$.

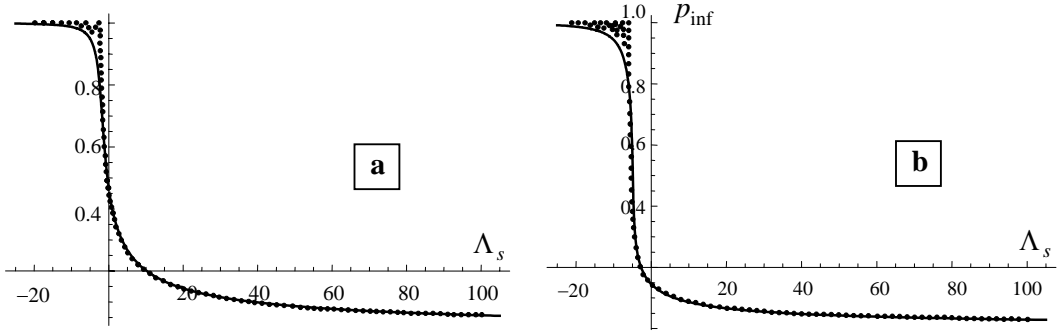

Fig. 3.2. Final transition probability $p_{inf}$ as a function of $\Lambda_s$ (solid lines - Eq. (9), filled circles – numerical result): (a) $\lambda = 0.2$ and (b) $\lambda = 0.05$.

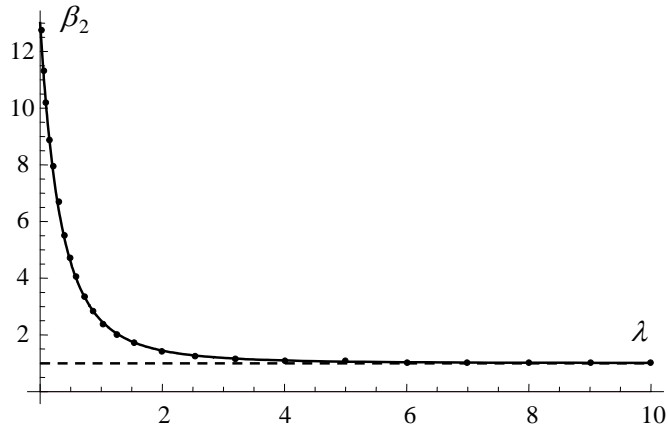

Fig. 4. The parameter $\beta_2$ as a function of $\lambda$.

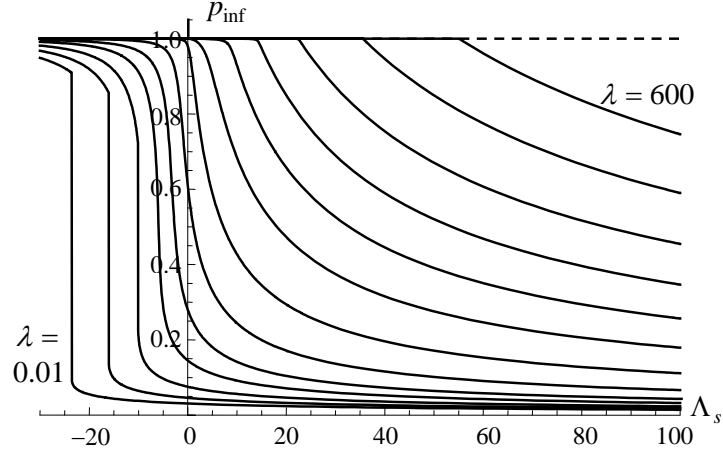

Fig. 5. From left to right: the dependence $p_{\inf}(\Lambda_s)$ defined by Eq. (9) for
$\lambda = 0.01, 0.015, 0.025, 0.05, 0.1, 0.3, 1.25, 5, 15, 40, 100, 250,$ and $600$.

Analyzing the curves shown on Fig. 5 we note that one may distinguish two interaction regimes depending on whether the second derivative $d^2 p_{\inf}/d\Lambda_s^2$ is negative or positive. For the positive case we note that for the *adiabatic* limit $\lambda \gg 1$ we have $k \approx 1$ and $P_{LZ} \approx \beta_2 \approx 1$, hence, Eq. (9) is considerably simplified reducing to a cubic one from which we readily obtain, via simple iterations, the known result [22] for the asymptotic behavior of $p_{\inf}$ for $\Lambda_s \gg \lambda$:

$$p_{\inf} \sim \left(\frac{3\sqrt{\lambda}}{2\Lambda_s}\right)^{2/3} \left[\frac{5}{2} - \left(\frac{15\sqrt{\lambda}}{4\Lambda_s}\right)^{2/3}\right]^{2/3}. \qquad (11)$$

The comparison of this approximation with Eq (9) and the corresponding numerical result is shown in Figs. 3.1a and 3.1b. The formula provides accurate approximation for $\Lambda_s \gg \lambda$ already starting from $\lambda \approx 3$ (Fig. 3.1a).

The 3D plot of $p_{\inf}$ calculated by means of numerical solution of the initial set of equations (1) as a function of $\lambda$ and $\Lambda_s$ is shown in Fig. 6. The curves shown in Fig. 5 correspond to successive sections of the surface presented in this figure for different fixed $\lambda$. Figs. 3.1-3.2 that present particular sections for $\lambda = 10, 1, 0.2,$ and $0.05$, respectively, indicate that Eqs. (9)-(10) accurately describe the surface $p_{\inf}(\lambda, \Lambda_s)$ everywhere except a very narrow region at the upper-left corner of the 3D plot ($\Lambda_s \ll -1, \lambda \ll 1$), where, as already mentioned above, small-amplitude oscillations are observed in the numerical result.

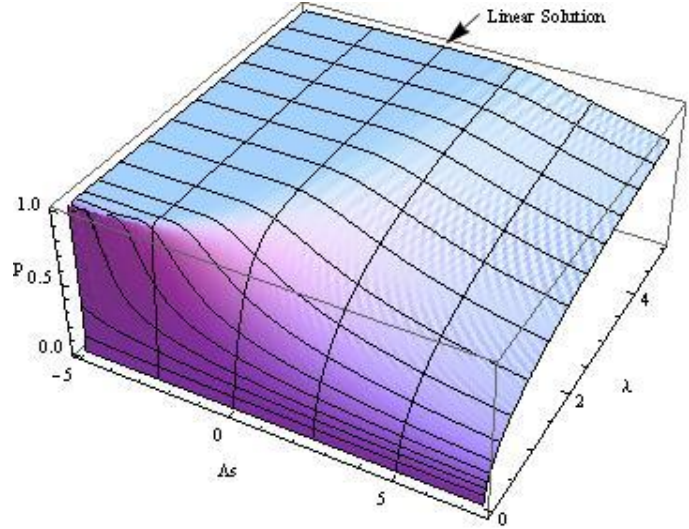

Fig. 6. The numerical result for the final transition probability $P = p_{\inf}(\lambda, \Lambda_s)$. The curve presenting the linear solution $p_{\inf}(\lambda, 0) \equiv P_{LZ} = 1 - e^{-\pi\lambda}$ is indicated by the arrow.

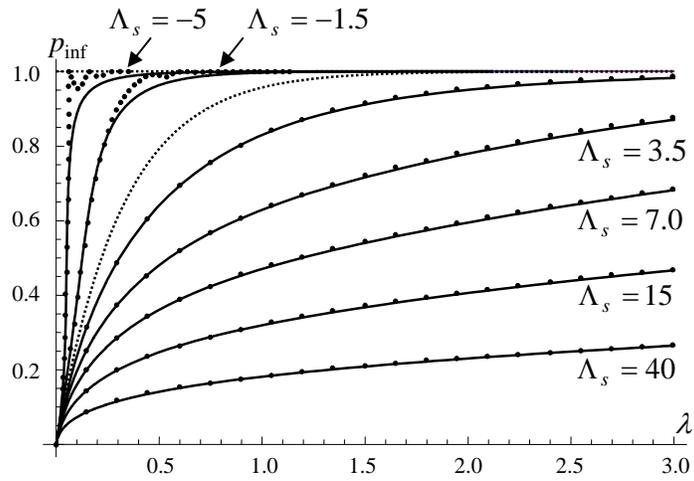

Fig. 7. The dependence $p_{\inf}(\lambda, \Lambda_s)$ for $\Lambda_s = -5, -1.5, 1.25, 3.5, 7, 15,$ and $40$. Solid lines - Eq. (9), filled circles – numerical result. The linear solution $P_{LZ}$ for $\Lambda_s = 0$ is reproduced exactly (dotted line).

Alternative sections of Fig. 6 for different fixed $\Lambda_s$ are shown in Fig. 7. We see that Eqs. (9)-(10) accurately describe both adiabatic ($\lambda \gg 1$) and sudden ($\lambda \ll 1$) regimes (note that for $\Lambda_s = 0$ Eq. (9) produces the exact linear solution $P_{LZ}$). For large negative $\Lambda_s$, the final transition probability undergoes abrupt, nearly vertical jump to $p_{\inf} = 1$ followed by oscillations which dump to the line $p_{\inf} = 1$ within a small intermediate region of small $\lambda$. While Eq. (9) well describes the initial region where the jump occurs, it does not describe the oscillations (we note that in this region our result approximately gives the lower bund of the numerical solution). I hope to address this point in a future study.

**Acknowledgments**

This research has been conducted within the scope of the International Associated Laboratory IRMAS. The work was supported by the Armenian National Science and Education Fund (ANSEF Grant No. 2010-MP-2186).